\documentclass[%
 reprint,
superscriptaddress,
 amsmath,amssymb,
 aps,
 prl
]{revtex4-2}

\usepackage{graphicx}
\usepackage{dcolumn}
\usepackage{bm}
\usepackage{color}
\usepackage{pdfpages} 
\usepackage{pgffor} 

\makeatletter
\AtBeginDocument{\let\LS@rot\@undefined}
\makeatother

\def\supplementfilename{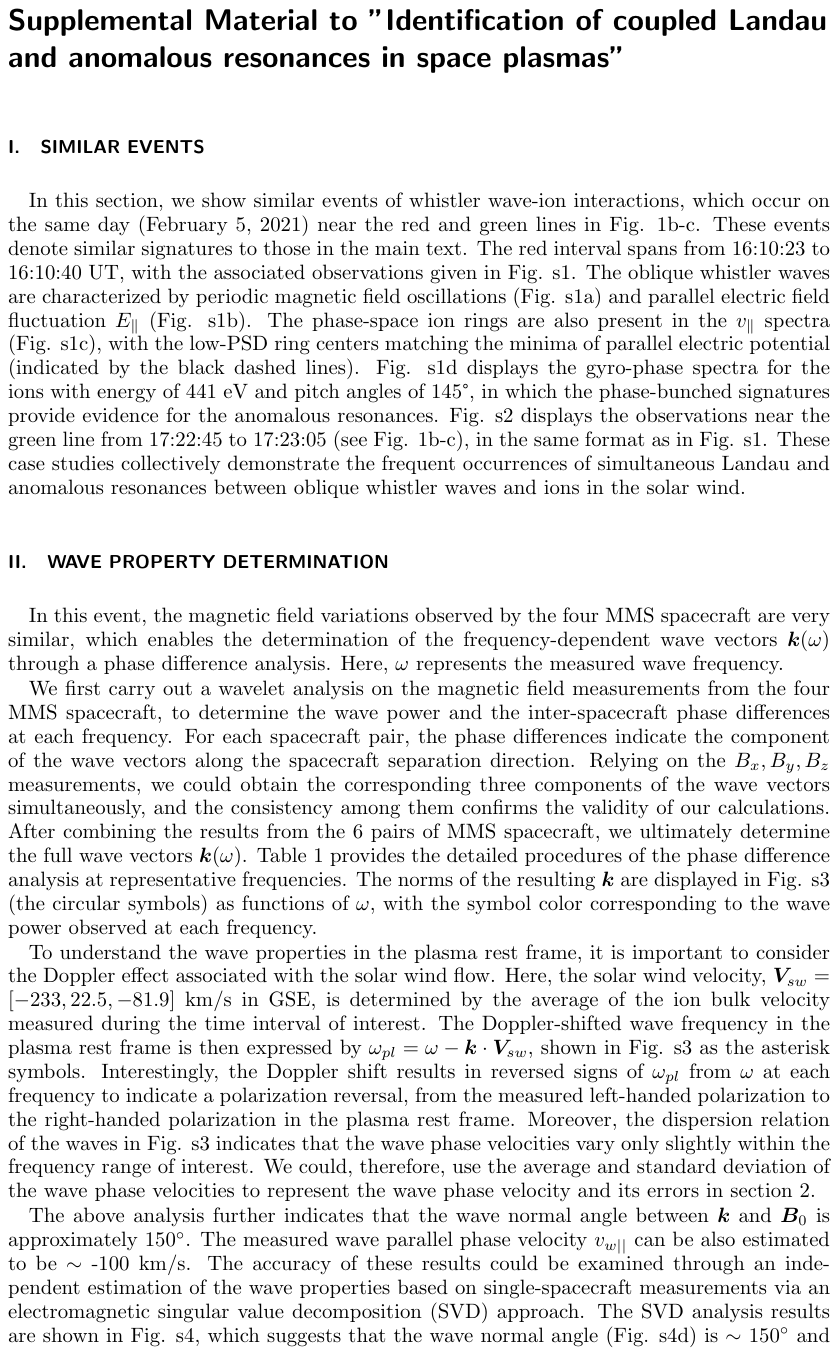}

\pdfximage{\supplementfilename}
\def\numbersupplementpages{\the\pdflastximagepages}

\newif\ifarXiv
\arXivtrue 
\begin{document}

\preprint{APS/123-QED}

\title{Identification of coupled Landau and anomalous resonances in space plasmas}

\author{Jing-Huan Li}
 
\author{Xu-Zhi Zhou}%
\email{xzzhou@pku.edu.cn}

\author{Zhi-Yang Liu}

\author{Shan Wang}
\affiliation{School of Earth and Space Sciences, Peking University, Beijing 100871, China}

\author{Anton V. Artemyev}
\affiliation{Institute of Geophysics and Planetary Physics, University of California, Los Angeles, California 90095, USA}

\author{Yoshiharu Omura}
\affiliation{Research Institute for Sustainable Humanosphere, Kyoto University, Kyoto 611-0011, Japan}

\author{Xiao-Jia Zhang}
\affiliation{Department of Physics, University of Texas at Dallas, Richardson, TX 75080, USA}

\author{Li Li}

\author{Chao Yue}

\author{Qiu-Gang Zong}
\affiliation{School of Earth and Space Sciences, Peking University, Beijing 100871, China}%

\author{Craig Pollock}
\affiliation{Denali Scientific, Fairbanks, AK 99709, USA}

\author{Guan Le}
\affiliation{NASA Goddard Space Flight Center, Greenbelt, MD 20771, USA}

\author{James L. Burch}
\affiliation{Southwest Research Institute, San Antonio, TX 78238, USA}

\date{\today}

\begin{abstract}
Wave-particle resonance, a ubiquitous process in the plasma universe, occurs when resonant particles observe a constant wave phase to enable 
sustained energy transfer. Here, we present spacecraft observations of simultaneous Landau and anomalous resonances between oblique whistler 
waves and the same group of protons, which are evidenced, respectively, by phase-space rings in parallel-velocity spectra and phase-bunched 
distributions in gyro-phase spectra. Our results indicate the coupling between Landau and anomalous resonances via the overlapping of the 
resonance islands.
\end{abstract}

\maketitle

\section{\label{sec:intro}1. Introduction}

In collisionless space plasma environments, the dynamics of charged particles are largely governed by their interactions with plasma waves. The wave-particle energy transfer becomes especially efficient when they resonate, namely, the particle observes a constant wave phase. The resonance condition\cite{nunn2015} in magnetized plasma is usually given by
\begin{eqnarray}
\omega-k_{\parallel}v_{\parallel}=n\Omega
\label{equ:one}
\end{eqnarray}
where $\omega$ is the wave frequency, $k_{\parallel}$ is the parallel wavenumber, $v_\parallel$ is the particle's parallel velocity, $\Omega$ is the particle's gyrofrequency, and $n$ is an integer. A fundamental resonant process, the Landau resonance, applies when $n=0$ so that the particle moves at the same speed as the wave propagation along the background magnetic field to observe a constant phase $\phi_{E\parallel}$ of the wave parallel electric field $E_\parallel$. Another important resonance is the first-order 
cyclotron resonance, which occurs at $n=1$ as the Doppler-shifted wave frequency matches the particle's gyrofrequency. Therefore, the phase difference $\zeta$ between the particle’s perpendicular velocity and the wave magnetic field remains constant\cite{Omura2021}. Accordingly, a resonant velocity\cite{Omura2021} is defined as 
\begin{eqnarray}
V_r = (\omega-\Omega)/k_\parallel 
\label{equ:two}
\end{eqnarray}
to represent the parallel velocity of the particle in cyclotron resonance with the waves. 

As a resonant particle gains or loses energy from the waves, its parallel velocity $v_\parallel$ also changes to deviate from the resonant velocity, which leads to the variation of wave-particle phase difference ($\phi_{E\parallel}$ for Landau resonance or $\zeta$ for cyclotron resonance). For some resonant particles, however, the $\phi_{E\parallel}$ or $\zeta$ variations are periodic to form closed particle trajectories in the $v_\parallel-\phi_{E\parallel}$ or $v_\parallel-\zeta$ phase space. This process is called nonlinear trapping, and the phase-space region occupied by the trapped particles is named resonance island\cite{anton2014,Omura2021,taoxin2020}. These nonlinear wave-particle interactions play an important role in the dynamics of space environments. In the radiation belts, they are responsible for the frequency chirping of chorus waves\cite{Omura2021,taoxin2020} and the acceleration/precipitation of relativistic electrons\cite{foster2017,zhangxiaojia2022}. In the foreshock solar wind, they produce gyrophase-bunched 
ions\cite{hoshino_simulation_bunching, mazelle_bunching_observation, romanelli_bunching_observation} and accelerate/thermalize electron beams\cite{anton_whistler_foreshock, shi_acceleration}.

The cyclotron resonance can be further modified into anomalous resonance if the particle's gyromotion around the background magnetic field is significantly perturbed by the wave field, which changes the particle's angular velocity and thus revises the resonance condition (1)\cite{albert2021,anton2021,Bortnik2022,kitahara2019,jinghuan2022}. Moreover, the original resonance island can be dissociated into two separate islands when anomalous resonance occurs\cite{jinghuan2022}. Interestingly, the Landau and anomalous/cyclotron resonances may occur simultaneously for oblique-propagating waves\cite{behar2020resonant}, although it remains unclear whether or how they are coupled.

Spacecraft identification of wave-particle resonance often relies on estimations of wave-particle energy transfer \cite{chen2019evidence, afshari2021importance, klein2020diagnosing} and/or averaged distributions of particles indicative of their diffusion processes \cite{he2015evidence,bowen2022situ,behar2020resonant,mcmanus2024proton}. Recently, the high-resolution particle observations from Magnetospheric Multiscale (MMS) spacecraft provide more detailed insights into particle dynamics within a single wave period. For instance, the observations of phase-bunched stripes in gyro-phase spectra\cite{kitamura2018, LZY2022} and V-shaped structures in pitch-angle spectra\cite{LZY2021} shed new light on nonlinear trapping of resonant particles in the wave field.

In this paper, we use MMS observations to study the interactions between large-amplitude oblique whistler waves in the terrestrial foreshock and solar wind protons. In the parallel direction, a series of proton phase-space rings are observed with a one-to-one correspondence to $E_\parallel$ variations indicative of nonlinear Landau resonance. In the perpendicular plane, the periodic appearance of phase-bunched ion structures supports the occurrence of anomalous resonance. This is a direct observational identification of simultaneous Landau and anomalous resonances, and their coupling demonstrates the important roles of oblique waves in governing the complicated particle dynamics in space.

\section{\label{sec:ob}2. Observations}

\subsection{\label{sec:overview}2.1 Overview}

At $\sim$ 1630 UT on February 5, 2021, the four-spacecraft MMS constellation was located near the terrestrial bow shock at [13.7, 6.1, -6.1 ] $R_E$ (Earth radius) in Geocentric Solar Eclipse (GSE) coordinates (with $\boldsymbol{x}$ axis towards the Sun and $\boldsymbol{z}$ axis perpendicular to the ecliptic plane). The onboard instruments offer high-resolution electromagnetic and particle measurements to analyze the wave-particle interactions \cite{fpi2016, mmsfield2016, MMSHPCA}.
In this event, the protons contribute over 99\% to the plasma density, and we hereinafter neglect the contribution of heavier ions.

Figure 1 provides a 1.5-hour overview of MMS3 observations. Similar features are also observed by other MMS satellites (not shown) due to their minor separations. The energy spectrum of ion energy fluxes in Fig. 1a indicates the back-and-forth crossings of the bow shock, characterized by the shifts between narrower and wider energy distributions that represent the foreshock solar wind and the thermalized magnetosheath ions, respectively. These features also agree with the higher and lower plasma bulk speeds in the foreshock and magnetosheath, respectively (see Fig. 1b). The magnetic field $B_y$ in Fig. 1c indicates the occurrence of ultralow-frequency (ULF) waves at the period of 30-50s, which have been commonly reported in these regions\cite{hoppe1981upstream, hoppe1983, wilson2016}.

Higher-frequency waves are also present in this event. Fig. 1d shows the magnetic measurements from a 3-min segment, which exhibit oscillations with a decreasing wave period from 10s to 4s in the shadowed interval. The waves are also characterized by its large amplitude $B_1$, around 1.5 times the background field $B_0$. The shadowed interval is then chosen for in-depth analysis of wave properties and wave-ion interactions. Similar findings can be also drawn from observations near the red- and green lines in Fig. 1b-c, as detailed in the Supplementary Material \cite{supple}.

\subsection{\label{sec:wave_property}2.2 Wave property}

\begin{figure}
\includegraphics{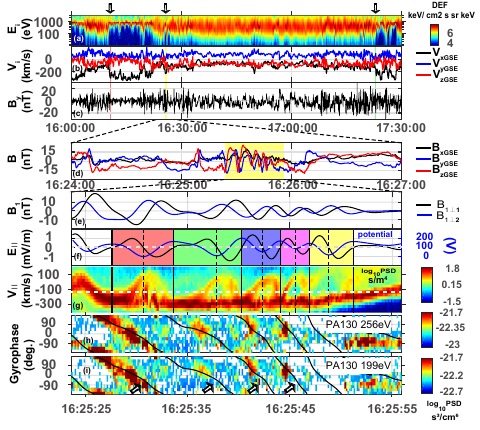}
\caption{\label{fig1} MMS3 observations of oblique whistler waves. The top panels provide a 1.5-h overview, including (a) ion energy spectrum, (b) bulk velocity, and (c) magnetic field $B_y$. (d) A 3-min overview of magnetic field variations. The zoomed-in view of the shadowed interval is shown in the bottom panels, including (e) $\boldsymbol{B}_{1\perp}$, 
(f) $E_\parallel$, overplotted by the integrated potential. (g) $v_\parallel-t$ spectra, (h-i) gyro-phase spectra. The black lines represent the gyro-phase of -$\boldsymbol{B}_1$.}
\end{figure}

We first determine, based on the four-spacecraft measurements of the magnetic field, the wave vectors at each frequency \cite{dudok1995determination}. The phase difference analysis \cite{dudok1995determination}, with details given in the Supplementary Material\cite{supple}, provides the measured wave dispersion relation in the reference frame moving with the spacecraft, which can be further transformed into the plasma rest frame. This result reveals that the observed wave phase velocity $\boldsymbol{v}_w$ varies only slightly within the wave frequency range of interest, which is approximately $[-84.8\pm4.1, 32.0\pm5.0, 4.1\pm4.0]$ km/s. The estimated wave normal angle, approximately $150^{\circ}$ between $\boldsymbol{v}_w$ and the background field $\boldsymbol{B}_0$, demonstrates an oblique wave propagation. The corresponding parallel wavenumbers $k_\parallel$ range from $\sim -0.6/d_p$ to $\sim -1.5/d_p$, where $d_p \sim 100$km is the proton inertial length. These estimations are further validated by the application of electromagnetic singular value decomposition analysis \cite{santolik2003singular}, which yields similar results (see Supplementary Materials \cite{supple} for details).

We further establish a field-aligned coordinate (FAC) with the parallel direction $\boldsymbol{e}_\parallel$ defined along $\boldsymbol{B}_0$, which is derived from a 15s running average of the field measurements. In the perpendicular plane, the $\boldsymbol{e}_{\perp1}$ direction is along the cross product of $\boldsymbol{B}_0$ and $\boldsymbol{v}_w$, 
and $\boldsymbol{e}_{\perp2}$ completes the triad. Based on this coordinate, the particle gyro-phase angle is defined by the angle between its perpendicular velocity $\boldsymbol{v}_\perp$ and the $\boldsymbol{e}_{\perp1}$ axis. The zoomed-in observations of the waves and the associated proton distributions, organized in the FAC coordinate, are shown in the bottom panels of Fig. 1. The perpendicular wave field $\boldsymbol{B}_{1\perp}$, filtered from 0.1 to 0.3 Hz (close to proton gyrofrequency $f_{cp} \sim 0.3$Hz), is displayed in Fig. 1e. Obviously, its $\perp_2$-component leads its $\perp_1$-component in phase by $\pi$/2, which indicates a left-handed polarization. In the plasma rest frame, however, the waves propagate sunward at the speed of $[121.5\pm5.4,-45.9\pm6.9,-5.9\pm5.9]$ km/s, which indicates a polarization reversal into right-handed waves due to the Doppler effect (with a frequency around -1.4 rad/s) \cite{LZY2021,wilson2016}. The waves are also associated with $E_\parallel$ oscillations (black line in Fig. 1f), and therefore, are identified as oblique whistler waves frequently reported in the foreshock \cite{wilson2016, hoppe1980whistler}. 

\subsection{\label{sec:level2}2.3 Wave-proton interactions}
To study the wave-proton interactions, we separate the time interval of interest into five wave periods (see the colored shadows in Fig. 1f) based on the electric-field phase angle $\phi_{E\parallel}$, which increases from 0° to 360° during each period. The dashed lines mark $\phi_{E\parallel}$=180°, where the parallel electric potential (blue line in Fig. 1f, integration of the observed $E_\parallel$) minimizes. 

Fig. 1g shows the ion phase-space density (PSD) distributions as functions of $v_\parallel$ and time, which display a series of rings with high PSDs surrounding their lower-PSD centers. All the ring centers are located near the potential minimum with $\phi_{E\parallel}$ $\sim$ 180° (the dashed lines), and their parallel velocities ($v_\parallel$ $\sim$ -100 km/s) are close to the parallel wave speed $v_{w\parallel}$ = -105.0 $\pm$ 4.3 km/s. These five rings, as will be shown in the simulation section, represent the phase-space 
trajectories of the ions trapped within the wave-carried potential wells (the regions surrounding the potential minima), in which the ion parallel velocities vary between -300 and 100 km/s. The velocity variations, $\pm200$ km/s in the wave rest frame, are consistent with a 200-V potential well (shown in Fig. 1f) and the width of the Landau resonance island\cite{nunn2015}. In other words, they are signatures of nonlinear Landau trapping for ions resonating with the whistler waves. 

Interestingly, the same ion population is also modulated in the perpendicular plane. Fig. 1h-i shows the gyro-phase spectra of the 256- and 199-eV ions at the pitch-angle range of 130°±20° (corresponding to a $v_\parallel$ range from -200km/s to -100 km/s), which are characterized by periodic occurrences of inclined stripes with enhanced PSDs. Also shown in Fig. 1h-i (as black lines) are the gyro-phase of -$\boldsymbol{B}_1$, which are mostly aligned with the inclined stripes to indicate the phase-bunched ion behavior at $\zeta$$\sim$180°. These features are usually treated as diagnostic signatures of cyclotron resonance\cite{kitamura2018,LZY2022}. However, based on the wave properties given above, the cyclotron resonance velocity $V_r$ is expected to be $\sim$ 108.7 $\pm$ 58.8 km/s, which differs dramatically from the parallel velocity of the phase-bunched ions.
\begin{figure}
\includegraphics{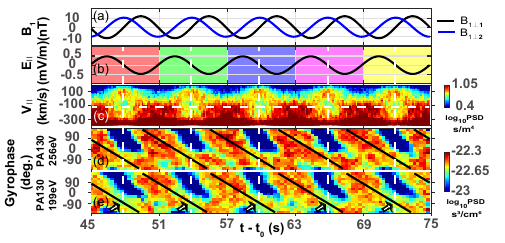}
\caption{\label{fig:2}Virtual spacecraft observations. Panels a-e are in the same format as Fig. 1e-1i.}
\end{figure}

The deviation from the classical cyclotron-resonance theory could be explained by the occurrence of anomalous resonance (also named anomalous trapping\cite{kitahara2019}, distinctly different from anomalous cyclotron resonance\cite{omura2012nonlinear}), which takes place when the wave-associated forces, including the $q\boldsymbol{v}_\parallel \times \boldsymbol{B}_1 $ and $q\boldsymbol{E}_1$ forces in the perpendicular plane, are comparable to the background $q\boldsymbol{v}_\perp \times \boldsymbol{B}_0$ force. This is made possible by the comparable $B_0$ and $B_1$ magnitudes. The additional, wave-associated forces modify the conventional gyrofrequency $\Omega$, and consequently, provide a new term in the classical condition (1)\cite{kitahara2019,jinghuan2022}. 
The resulting resonance speed (for ions at $\zeta$=180°) \cite{jinghuan2022} equals 
\begin{eqnarray}
V_r^{'} = V_r - \frac{V_r-v_w}{k_\parallel v_\perp/\Omega_1 +1}
\label{equ:three}
\end{eqnarray}
where $V_r$ is obtained from equation (2), and $\Omega_1=B_1 q/m$ is the nominal gyrofrequency associated with $B_1$. Based on equation (3), $V_r^{'}$ is estimated at approximately $-200$ km/s, corresponding to the pitch angle of 130°±20° for 100-500eV ions. Therefore, we speculate that the phase-bunched ion distributions in Fig. 1h-i are caused by anomalous, rather than classical cyclotron, resonance. The observations suggest that oblique whistler waves could interact with the ions via Landau and anomalous resonances simultaneously, which will be examined through test-particle simulations.

\section{\label{simu}3. Simulation}
We next follow the test-particle approach of ref.\cite{jinghuan2022} to study the interaction between oblique whistler waves and solar wind ions. In the simulation, the ion velocity distributions at any time and location are determined by tracing the ions backward in time, since Liouville’s theorem requires PSD conservation along particle trajectories (see Supplementary Material for detailed simulation setup\cite{supple}). The 3D distributions are integrated to obtain the $v_\parallel$-t and/or gyro-phase spectrograms in the same format as observations in Fig. 1g-i. 

\begin{figure}
\includegraphics{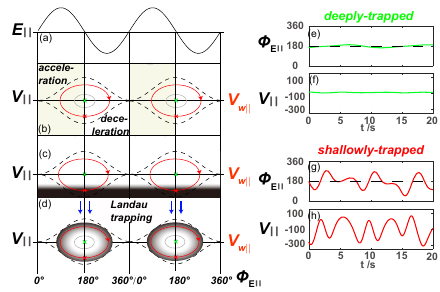}
\caption{\label{fig:3}Illustration of phase-space ring formation. (a) $E_\parallel$. (b) $v_\parallel$ (c) Initial $v_\parallel$ distribution, 
with black color denoting high PSDs. (d) Phase-mixing process. (e-h) $\phi_{E\parallel}$ and $v_\parallel$ variations for the green and red ions, respectively.}
\end{figure}

Fig. 2 shows virtual spacecraft observations of the simulated field and ion distributions since $t=t_0$+45s. Fig. 2a-b 
display the perpendicular wave magnetic field $\boldsymbol{B}_{1\perp}$ and parallel electric field $E_\parallel$, respectively. Fig. 2c shows the simulated $v_\parallel$-t spectrogram of the ion PSDs, which resembles the periodic appearance of phase-space rings with a one-to-one correspondence to $E_\parallel$ variations in Fig. 2b. The similarities between simulations and observations (compare Fig. 2c to Fig. 1g) enable us to understand the processes underlying the phase-space rings via analysis of the ion trajectories. 

A schematic framework of our analysis is shown in Fig. 3. Here, $E_\parallel$ profile is given in Fig. 3a, with its phase angle $\phi_{E\parallel}$ increasing from 0° to 360° during each wave period. The corresponding electric potential minimizes at $\phi_{E\parallel}$ = 180°, where the resonant 
ions are trapped nonlinearly (Landau trapping, see Fig. 3b for illustrations of closed trajectories in the $\phi_{E\parallel}$-$v_\parallel$ phase space). The green ion in Fig. 3b is phase-locked at the center of the potential well ($\phi_{E\parallel}$ = 180°), which represents the deeply-trapped ions with negligible $\phi_{E\parallel}$ and $v_\parallel$ variations. The wave-carried potential well can also trap the red ion in Fig. 3b despite larger variations in $\phi_{E\parallel}$ and $v_\parallel$. Since its trajectory is close to the separatrix of the resonance island, the ion belongs to the shallowly-trapped population. 

The initial PSDs for deeply- and shallowly-trapped ions are distinctly different, however. The initial ion distributions in \cite{supple} indicate PSD peaks at $v_\parallel\sim-370$ km/s (the background solar wind), which is much lower than $v_{w\parallel}$ and thus is located near the bottom of the $\phi_{E\parallel}$-$v_\parallel$ diagram in Fig. 3c (the black region, which represents the background ion concentration). Since the trajectories of the deeply-trapped ions (including the green ion) cannot intersect the high-PSD region, the phase space occupied by these ions (the resonance island center) can only have lower densities due to the PSD conservation. The shallowly-trapped ions, however, have trajectories intersecting the black region (the red trajectories) so that the high-PSD solar wind ions can easily access the outer edge of the resonance islands. Note that there are also regions with low initial PSDs along the red trajectories, and therefore, it would take longer than a Landau trapping period for the ions with higher and lower PSDs to mix in phase space and produce rings with intermediately-high PSDs (Fig. 3d). Here, the Landau-trapping period of $\sim$ 6s is indeed shorter than the formation time ($\sim$30s) of phase-space rings in the simulations. 

To examine the scenario of phase-space ring formation via Landau trapping, we launch two typical ions for their trajectories in the modeled field. The green ion trajectory in Fig. 3e-f shows minor oscillations of $\phi_{E\parallel}$ and $v_\parallel$ around 180° and -100 km/s (approximately the parallel wave velocity $v_{w\parallel}$), respectively. Their minor variations indicate that the ion belongs to the deeply-trapped population. On the other hand, the red ion experiences a larger variation of $\phi_{E\parallel}$ from $\sim$90° to $\sim$270° (Fig. 3g), which indicates that 
it is trapped, albeit more shallowly, within the potential well (otherwise, $\phi_{E\parallel}$ should cover the full range from 0° to 360°). Moreover, its $v_\parallel$ changes significantly from $\sim$-300km/s to $\sim$100km/s (Fig. 3h), which indicates that solar-wind ions with higher PSDs can indeed be transported in phase space along closed trajectories around the resonance island. Therefore, we conclude that the formation of phase-space ion rings is associated with particle trapping, serving as the observational identification of nonlinear Landau resonance.

We next return to the simulated ion gyro-phase spectra (Fig. 2d-e), which reproduce the phase-bunched features in observations (Fig. 1h-i). To confirm that the phase-bunched features are caused by anomalous resonance, we show in Fig. 4a the $v_\parallel-\zeta$ portrait of the ion phase-space trajectories. Here, for simplicity, we assume field-aligned wave propagation, with the effect of oblique propagation to be discussed later on. There are two resonance islands occupying different $v_\parallel$ ranges, centered at $\zeta$=0° (purple lines) and $\zeta$=180° (blue lines), respectively\cite{jinghuan2022}. The blue island centered at $\zeta$=180° corresponds to lower $v_\parallel$ values. Therefore, the nearby ion trajectories, including the closed (blue) and the traversing trajectories (yellow), can access the background solar-wind ions ($v_\parallel\sim-370$km/s), which leads to the concentration of high-PSD ions around the $\zeta$=180° island. Fig. 4a also displays the observed $v_\parallel-\zeta$ spectrum (averaged from 1625:33.7 to 1625:35.7), in which the high-PSD ions are indeed concentrated around $\zeta$=180°. This scenario is consistent with the superimposed trajectories, contributing to the inclined stripes in observations (Fig. 1h-i) and simulations (Fig. 2d-e).
\begin{figure}
\includegraphics{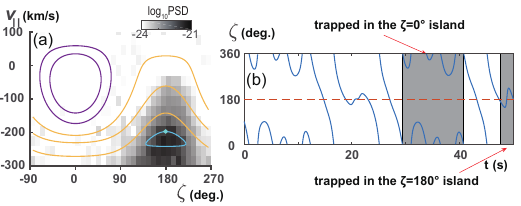}
\caption{\label{fig4} Typical particle trajectories in the simulation. (a) The $v_\parallel-\zeta$ spectrum, overplotted by ion trajectories. (b) $\zeta$ variation.}
\end{figure}

\section{\label{D_S}4. Discussion and Summary}
The simulated ion trajectories in oblique waves could be more complicated than expected from the $v_\parallel-\zeta$ diagram. Fig. 4b shows the $\zeta$ variations of a sample ion, which switches back and forth between traversing (when $\zeta$ varies monotonically from 360° to 0°) and trapped trajectories (when $\zeta$ oscillates around $\zeta$=0° or 180°) within either island. The switch between different types of trajectories is usually attributed to field variations\cite{jinghuan2022, Omura2021}, but this is not the case here. Instead, the reason lies in the existence of $E_\parallel$ (a property of oblique whistler waves), which perturbs the parallel ion motion in the $v_\parallel-\zeta$ diagram and consequently destabilizes the ion trajectories. As a result, the phase-bunched stripes in Fig. 1i and 2e are not homogeneous, with minimum PSDs near $\phi_{E\parallel} \sim$180° (see arrows in Fig. 1i and 2e) where the ions are Landau trapped by $E_\parallel$ (see the white dashed lines in Fig. 2c, corresponding to the low-PSD centers).

In other words, the Landau and anomalous resonances are largely coupled. This coupling process, in analogy to the coupling between Landau and cyclotron resonances \cite{theory_overlapping, theory_overlapping_2}, is caused by the coexistence of the $q\boldsymbol{E}_\parallel$ and $q\boldsymbol{V}_\perp\times \boldsymbol{B}_\perp$ forces in the parallel direction, and more importantly, the overlapped resonance islands in the $v_\parallel$ range between -300km/s and 100km/s. The coupling does not only affect the anomalous resonance; the Landau resonance can also be perturbed by 
the $q\boldsymbol{v}_\perp\times \boldsymbol{B}_\perp$ force, although in our case, the $q\boldsymbol{v}_\perp\times \boldsymbol{B}_\perp$ force is much smaller than $q\boldsymbol{E}_\parallel$ due to the ion concentration at $\zeta$=180°. Consequently, the disturbances in Fig. 1g and Fig. 2c are much weaker. This additional force, however, could still revise the particle trapping period to expedite the phase-mixing process in the ring formation \cite{nunn2015}.

In summary, we demonstrate the simultaneous occurrence of Landau and anomalous resonances between oblique whistler waves and solar wind ions. The two resonant processes are strongly coupled through the overlapped resonance islands, which is made possible by the coexistence of wave-associated electric and Lorentz forces in the parallel direction. For anomalous resonance, the ion trapping and traversing behavior are perturbed by the additional electric force, which obscures the characteristic, phase-bunched features. The diagnostic signatures for Landau resonance (phase-space rings in Fig. 1g and 2c), however, are perturbed only slightly by the wave-associated Lorentz force due to the ion concentration near $\zeta$=180°. As the coupling strengthens, more chaotic ion motion and more complicated features could be expected, which may further revise the picture of foreshock wave-particle energy transfer. This is beyond the scope of this paper, and we will carry out a more systematic study to better understand the coupling process. 

\begin{acknowledgments}
This work was supported by the National Natural Science Foundation of China grants 42174184. 
We are grateful to the MMS mission for the high-resolution measurements. The MMS data are processed 
and analyzed using the IRFU-Matlab package.
\end{acknowledgments}


%

\ifarXiv
    

\section*{Supplementary material}

# Supplemental Material to "Identification of coupled Landau and anomalous resonances in space plasmas"

## I. SIMILAR EVENTS

In this section, we show similar events of whistler wave-ion interactions, which occur on the same day (February 5, 2021) near the red and green lines in Fig. 1b-c. These events denote similar signatures to those in the main text. The red interval spans from 16:10:23 to 16:10:40 UT, with the associated observations given in Fig. s1. The oblique whistler waves are characterized by periodic magnetic field oscillations (Fig. s1a) and parallel electric field fluctuation $E_\parallel$ (Fig. s1b). The phase-space ion rings are also present in the $v_\parallel$ spectra (Fig. s1c), with the low-PSD ring centers matching the minima of parallel electric potential (indicated by the black dashed lines). Fig. s1d displays the gyro-phase spectra for the ions with energy of 441 eV and pitch angles of 145°, in which the phase-bunched signatures provide evidence for the anomalous resonances. Fig. s2 displays the observations near the green line from 17:22:45 to 17:23:05 (see Fig. 1b-c), in the same format as in Fig. s1. These case studies collectively demonstrate the frequent occurrences of simultaneous Landau and anomalous resonances between oblique whistler waves and ions in the solar wind.

## II. WAVE PROPERTY DETERMINATION

In this event, the magnetic field variations observed by the four MMS spacecraft are very similar, which enables the determination of the frequency-dependent wave vectors $\boldsymbol{k}(\omega)$ through a phase difference analysis. Here, $\omega$ represents the measured wave frequency.

We first carry out a wavelet analysis on the magnetic field measurements from the four MMS spacecraft, to determine the wave power and the inter-spacecraft phase differences at each frequency. For each spacecraft pair, the phase differences indicate the component of the wave vectors along the spacecraft separation direction. Relying on the $B_x, B_y, B_z$ measurements, we could obtain the corresponding three components of the wave vectors simultaneously, and the consistency among them confirms the validity of our calculations. After combining the results from the 6 pairs of MMS spacecraft, we ultimately determine the full wave vectors $\boldsymbol{k}(\omega)$. Table 1 provides the detailed procedures of the phase difference analysis at representative frequencies. The norms of the resulting $\boldsymbol{k}$ are displayed in Fig. s3 (the circular symbols) as functions of $\omega$, with the symbol color corresponding to the wave power observed at each frequency.

To understand the wave properties in the plasma rest frame, it is important to consider the Doppler effect associated with the solar wind flow. Here, the solar wind velocity, $\boldsymbol{V}_{sw} = [-233, 22.5, -81.9]$ km/s in GSE, is determined by the average of the ion bulk velocity measured during the time interval of interest. The Doppler-shifted wave frequency in the plasma rest frame is then expressed by $\omega_{pl} = \omega - \boldsymbol{k} \cdot \boldsymbol{V}_{sw}$, shown in Fig. s3 as the asterisk symbols. Interestingly, the Doppler shift results in reversed signs of $\omega_{pl}$ from $\omega$ at each frequency to indicate a polarization reversal, from the measured left-handed polarization to the right-handed polarization in the plasma rest frame. Moreover, the dispersion relation of the waves in Fig. s3 indicates that the wave phase velocities vary only slightly within the frequency range of interest. We could, therefore, use the average and standard deviation of the wave phase velocities to represent the wave phase velocity and its errors in section 2.

The above analysis further indicates that the wave normal angle between $\boldsymbol{k}$ and $\boldsymbol{B}_0$ is approximately 150°. The measured wave parallel phase velocity $v_{w\parallel}$ can be also estimated to be $\sim$ -100 km/s. The accuracy of these results could be examined through an independent estimation of the wave properties based on single-spacecraft measurements via an electromagnetic singular value decomposition (SVD) approach. The SVD analysis results are shown in Fig. s4, which suggests that the wave normal angle (Fig. s4d) is $\sim$ 150° and

the wave parallel velocity (Fig. s4e) is $\sim$ -100 km/s at frequencies with significant wave power. The consistency between them confirms the reliability of both approaches and the accuracy of the estimated wave properties.

### III. SIMULATION SETUP

In the simulation, the ion velocity distributions at any time and location are determined by tracing the ions backward in time towards $t = t_0$, since Liouville's theorem requires the PSD conservation along particle trajectories. After assuming the initial ion distributions (in shifted-Maxwellian form, see equation (1)) and the electromagnetic field variations (see equations (2-3)), we trace a series of ions backward from the location of an immobile virtual spacecraft (located at the coordinate origin) to $t = t_0$, so that the 3D velocity distributions can be determined as functions of time.

Here, the initial ion distributions are

$$f = n_0 (\frac{m}{2\pi T})^{\frac{3}{2}} \exp\left(\frac{-m(\boldsymbol{v} - \boldsymbol{V_i})^2}{2T}\right), \tag{1}$$

where the bulk velocity $\boldsymbol{V_i} = (V_{i\perp 1}, V_{i\perp 2}, V_{i\parallel}) = (0,0,-370)$ km/s is obtained from the MMS spacecraft measurements (averaged from 14:10 to 14:20 UT when the solar wind is stable). The number density $n_0 = 6.4 \text{cm}^{-3}$ and temperature T=100eV are determined by the measurements during the shadowed interval.

We also assume a uniform background magnetic field $B_0$=9nT, superposed by the oblique-propagating wave fields,

$$\boldsymbol{E_1} = E_w[-\cos\varphi\, \boldsymbol{e}_{\perp_1},\ -\cos\theta\sin\varphi\, \boldsymbol{e}_{\perp_2},\ \sin\theta\sin\varphi\, \boldsymbol{e}_{\parallel}], \tag{2}$$

$$\boldsymbol{B_1} = B_w[-\sin\varphi\, \boldsymbol{e}_{\perp_1},\ \cos\theta\cos\varphi\, \boldsymbol{e}_{\perp_2},\ -\sin\theta\cos\varphi\, \boldsymbol{e}_{\parallel}], \tag{3}$$

where $\varphi = k_{\perp 2}r_{\perp 2} + k_\parallel r_\parallel - \omega t$ represents the wave phase, $B_w$ =12nT is the wave magnetic amplitude, $E_w$ =0.96 mV/m is the wave electric amplitude, and $\theta$=150° is the wave normal angle between $B_0$ and the wave vector $\boldsymbol{k} = (0, k_{\perp 2}, k_\parallel)$. Note that for simplicity, the waves are assumed to be monochromatic ($\omega$=1.05 rad/s) with a fixed wave phase speed $v_w = \omega/|\boldsymbol{k}| = 80$km/s in the spacecraft rest frame, which approximately matches the $v_\parallel$ values of the low-PSD ring centers in Fig. 1g, although the wave frequency gradually increases with time (see Fig. 1e) in real observations. This assumption is justified by the nearly constant $v_w$ in the observations for waves with different frequencies.

| the phase difference analysis for 0.13Hz waves | | | |
|---|---|---|---|
| components of $\boldsymbol{k}$ ($10^{-5}m^{-1}$) | from $B_x$ | from $B_y$ | from $B_z$ |
| along SC 1-2 | -0.78, 0.1, -0.02 | -0.9, 0.1, -0.02 | -0.91, 0.12, -0.02 |
| along SC 1-3 | -0.52, -0.15, 0.1 | -0.72, -0.21, 0.14 | -0.73, -0.21, 0.15 |
| along SC 1-4 | -0.39, 0.24, -0.49 | -0.34, 0.21, -0.43 | -0.32, 0.19, -0.41 |
| along SC 2-3 | -0.42, 0.6, -0.3 | -0.35, 0.5, -0.25 | -0.35, 0.5, -0.25 |
| along SC 2-4 | -0.78, 0.17, -0.19 | -0.84, 0.18, -0.21 | -0.84, 0.18, -0.21 |
| along SC 3-4 | -0.7, -0.-6, -0.09 | -0.84, -0.07, -0.11 | -0.83, -0.07, -0.11 |
| full $\boldsymbol{k}$ ($10^{-5}m^{-1}$) | -0.89, 0.39, 0.02 | | |
| the phase difference analysis for 0.17Hz waves | | | |
| components of $\boldsymbol{k}$ ($10^{-5}m^{-1}$) | from $B_x$ | from $B_y$ | from $B_z$ |
| along SC 1-2 | -1.08, 0.14, -0.02 | -1.06, 0.13, -0.02 | -1.01, 0.13, -0.02 |
| along SC 1-3 | -0.85, -0.24, 0.17 | -0.9, -0.26, 0.18 | -0.81, -0.23, 0.16 |
| along SC 1-4 | -0.41, 0.25, -0.52 | -0.37, 0.23, -0.47 | -0.32, 0.2, -0.41 |
| along SC 2-3 | -0.41, 0.59, -0.29 | -0.37, 0.53, -0.27 | -0.39, 0.57, -0.28 |
| along SC 2-4 | -1.02, 0.22, -0.25 | -0.98, 0.21, -0.24 | -0.92, 0.2, -0.23 |
| along SC 3-4 | -0.99, -0.08, -0.13 | -1.01, -0.08, -0.14 | -0.9, -0.07, -0.12 |
| full $\boldsymbol{k}$ ($10^{-5}m^{-1}$) | -1.05, 0.33, 0.09 | | |
| the phase difference analysis for 0.20Hz waves | | | |
| components of $\boldsymbol{k}$ ($10^{-5}m^{-1}$) | from $B_x$ | from $B_y$ | from $B_z$ |
| along SC 1-2 | -1.28, 0.16, -0.02 | -1.2, 0.15, -0.02 | -1.04, 0.13, -0.02 |
| along SC 1-3 | -1, -0.28, 0.2 | -1.07, -0.3, 0.2 | -0.85, -0.24, 0.17 |
| along SC 1-4 | -0.59, 0.36, -0.75 | -0.38, 0.23, -0.49 | -0.31, 0.19, -0.39 |
| along SC 2-3 | -0.5, 0.72, -0.36 | -0.39, 0.56, -0.28 | -0.4, 0.58, -0.29 |
| along SC 2-4 | -1.27, 0.27, -0.31 | -1.1, 0.24, -0.27 | -0.94, 0.2, -0.23 |
| along SC 3-4 | -1.23, -0.1, -0.17 | -1.16, -0.1, -0.16 | -0.93, -0.08, -0.12 |
| full $\boldsymbol{k}$ ($10^{-5}m^{-1}$) | -1.28, 0.45, 0.07 | | |

TABLE I. Frequency-dependent wave vectors determined through the phase difference analysis. For each spacecraft pair, the components of $k$ along the direction of their separation are determined from $B_x, B_y, B_z$ measurements. The full $k$ is then determined through a best fit procedure.

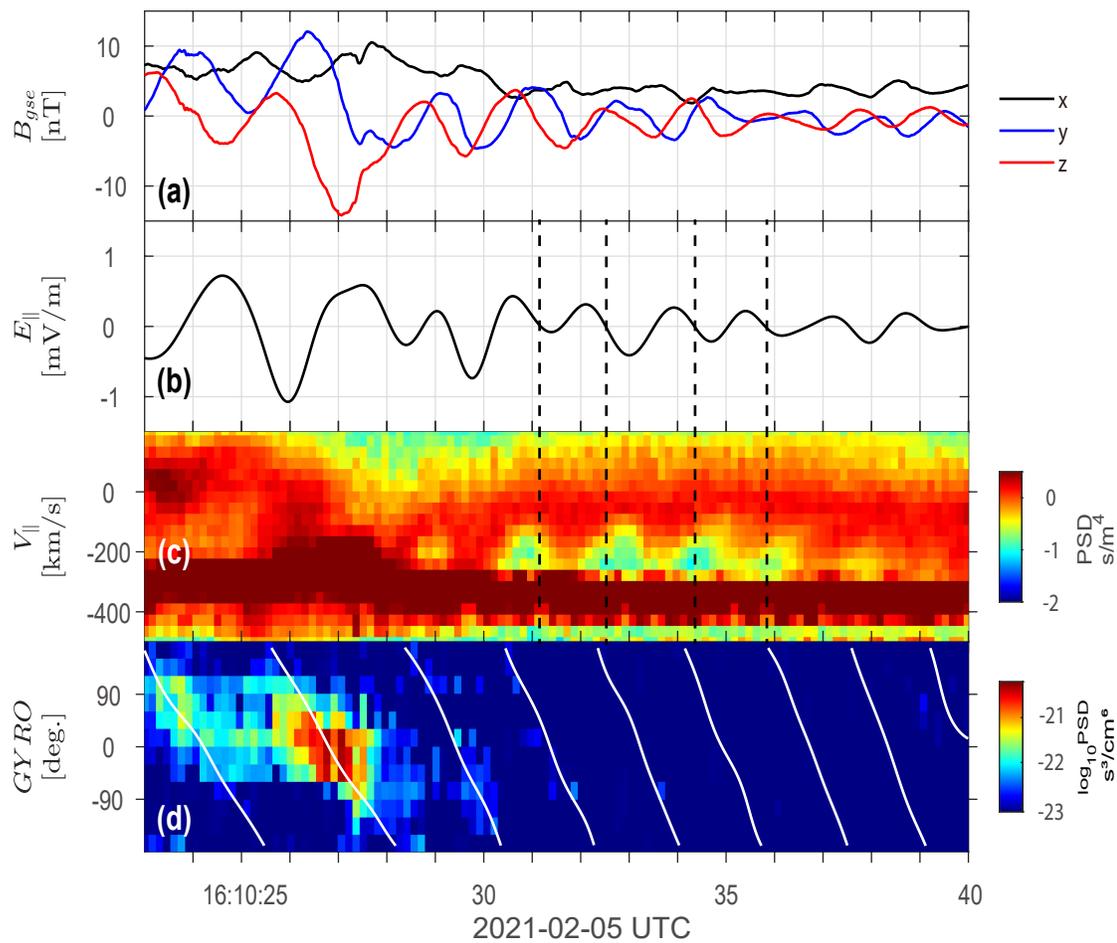

FIG. s1. MMS3 observations of oblique whistler waves during the red-shadowed interval in Fig. 1b-c. (a) magnetic field. (b) parallel wave electric field. (c) parallel velocity $v_\parallel$ spectra. (d) gyro-phase spectra, with the white lines representing the gyro-phase of $-\boldsymbol{B}_1$.

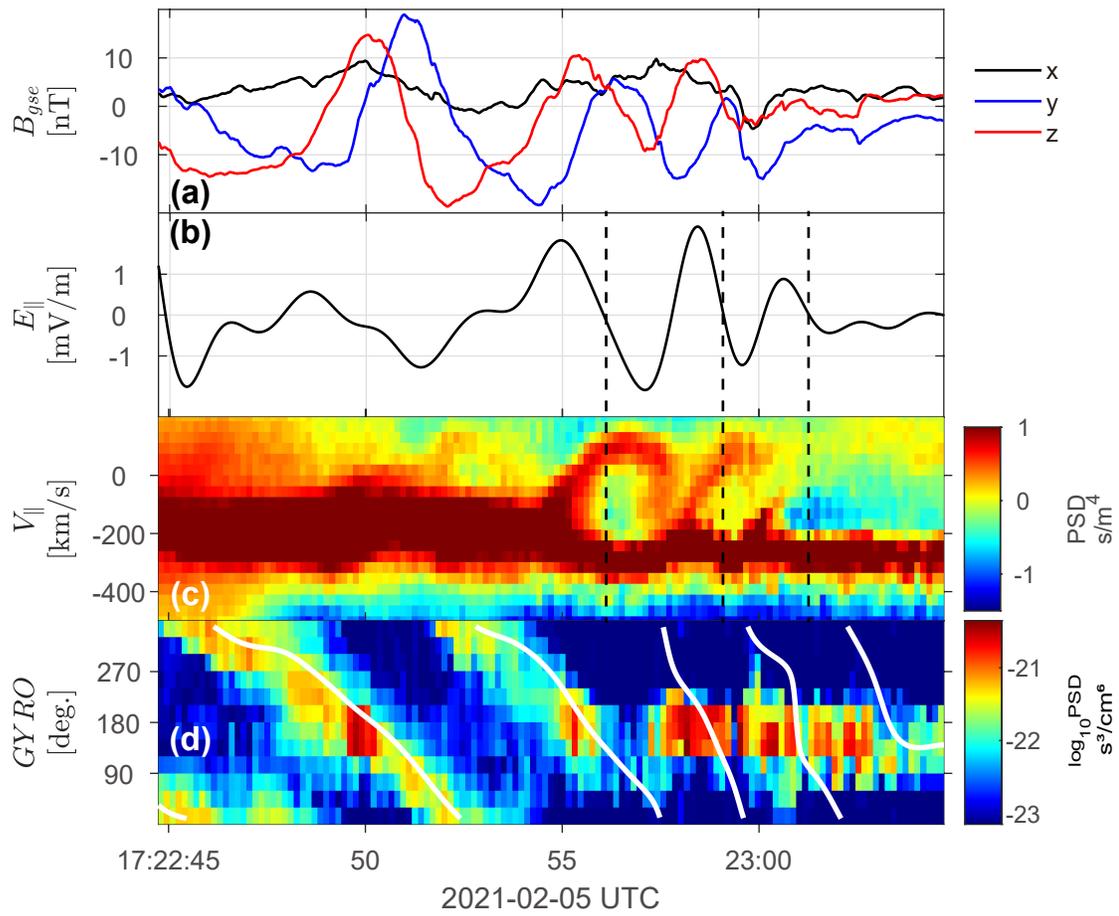

FIG. s2. MMS3 observations of oblique whistler waves during the green-shadowed interval in Fig. 1b-c, in the same format as in Fig. s1.

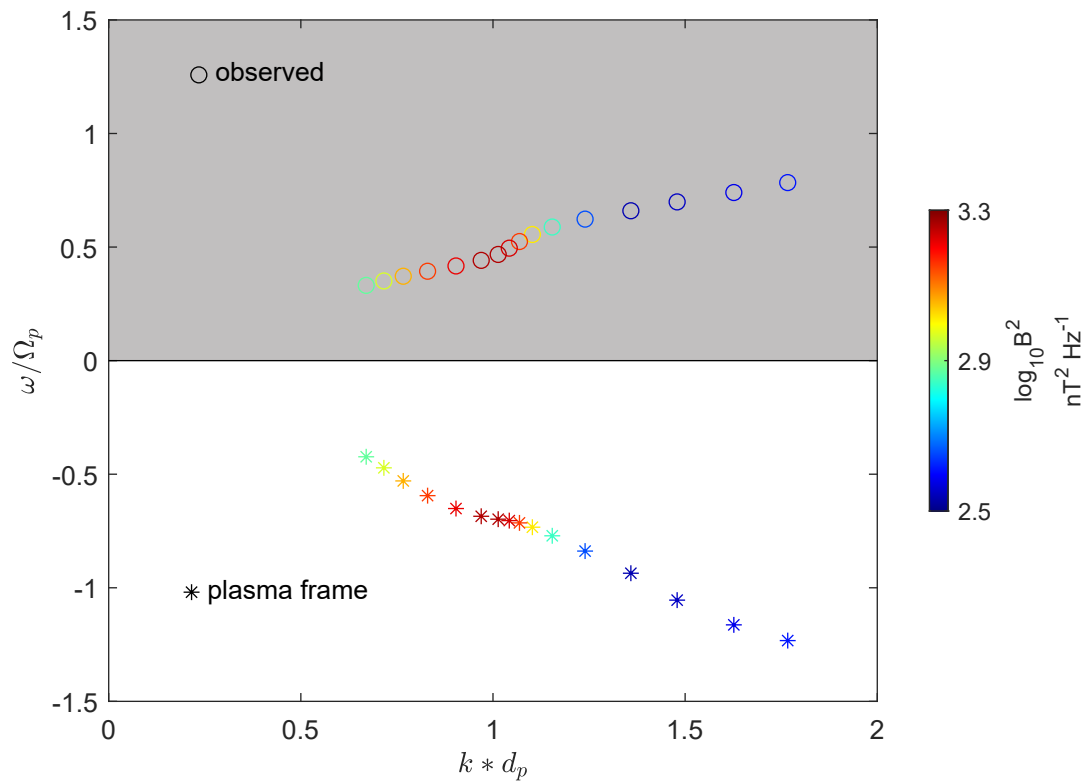

FIG. s3. The $\omega$-$k$ scatter plot for the observed waves, determined via the multi-spacecraft phase difference analysis. The vertical axis represents the wave frequency, normalized by the proton gyrofrequency $\Omega_p \sim 1.9$ rad/s, in the reference frame moving with spacecraft (circles) and plasma flow (asterisks), respectively. The horizontal axis represents the norm of the wave vector, normalized by the proton inertial length $d_p \sim 100$ km. The color indicates the observed wave power at the corresponding frequency.

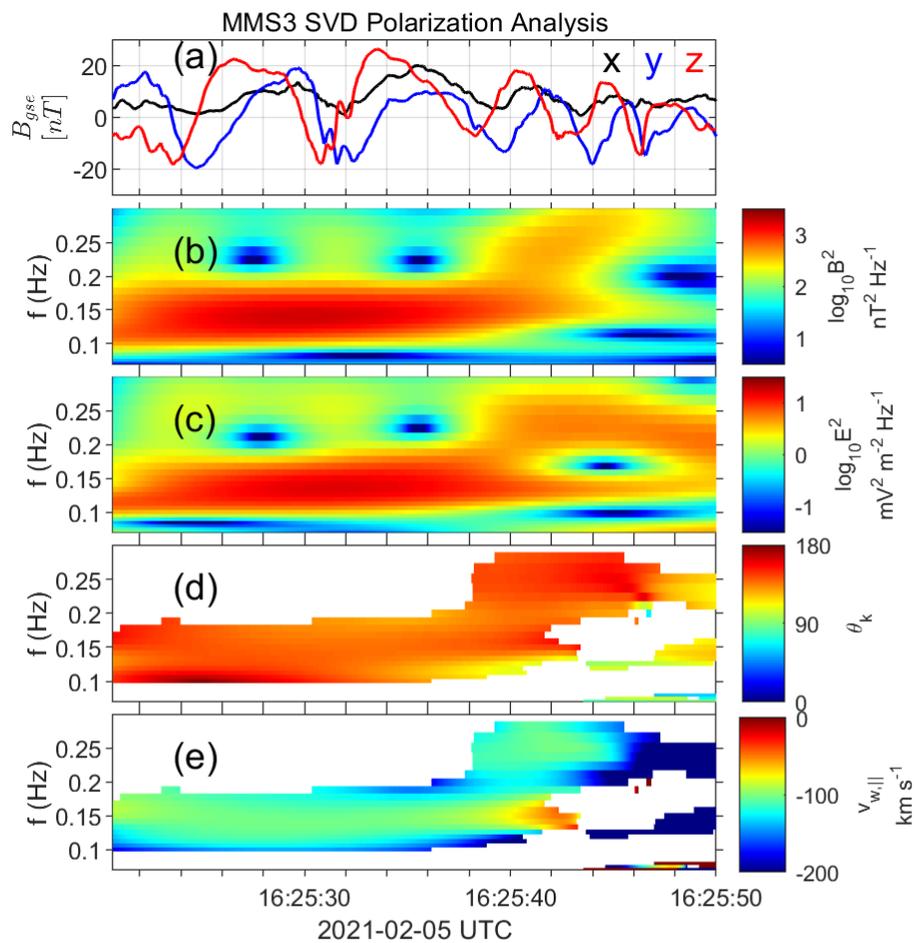

FIG. s4. The MMS3 observations and analysis results from the electromagnetic SVD technique. (a) Magnetic field in GSE coordinates. (b and c) Wavelet power spectral density of the magnetic field and electric field, respectively. (d and e) The wave normal angle and the parallel wave phase velocity determined via the electromagnetic SVD method, respectively. Only the waves with power greater than 2 $mV^2 m^{-2} Hz^{-1}$ are shown.


\fi

\end{document}